\documentclass[acmsmall,manuscript]{acmart}
\usepackage[noend]{algpseudocode}

\usepackage[ruled,lined,noend]{algorithm2e}
\SetKwRepeat{Do}{do}{while}
\makeatletter
\let\oldnl\nl
\newcommand{\nonl}{\renewcommand{\nl}{\let\nl\oldnl}}
\makeatother
\newlength\mylen
\newcommand\myinput[1]{%
  \settowidth\mylen{\KwIn{}}%
  \setlength\hangindent{\mylen}%
  \hspace*{\mylen}#1\\}

\usepackage{bm}

\DeclareMathOperator*{\argmax}{arg\,max}
\DeclareMathOperator*{\argmin}{arg\,min}

\newcommand{\norm}[1]{\left|\left|#1\right|\right|}

\AtBeginDocument{%
  }

\copyrightyear{2023}
\acmYear{2023}
\setcopyright{acmlicensed}\acmConference[RecSys '23]{Seventeenth ACM Conference on Recommender Systems}{September 18--22, 2023}{Singapore, Singapore}
\acmBooktitle{Seventeenth ACM Conference on Recommender Systems (RecSys '23), September 18--22, 2023, Singapore, Singapore}
\acmPrice{15.00}
\acmDOI{10.1145/3604915.3608851}
\acmISBN{979-8-4007-0241-9/23/09}

\begin{document}

\title{Topic-Level Bayesian Surprise and Serendipity for Recommender Systems}

\author{Tonmoy Hasan}
\authornote{Hasan contributed mainly the dataset processing, the manual annotation, the implementation, tuning, and evaluation of the proposed methods. Bunescu contributed mainly the task formulations, the algorithms, the formal definitions of surprise and serendipity, the overall study design, and the writing.}
\email{thasan1@uncc.edu}
\author{Razvan Bunescu}
\authornotemark[1]
\email{razvan.bunescu@uncc.edu}
\orcid{1234-5678-9012}
\affiliation{%
  \institution{University of North Carolina at Charlotte}
  \streetaddress{9201 University City Blvd}
  \city{Charlotte}
  \state{North Carolina}
  \country{USA}
  \postcode{28223}
}

\renewcommand{\shortauthors}{Hasan and Bunescu}

\begin{abstract}
A recommender system that optimizes its recommendations solely to fit a user's history of ratings for consumed items can create a filter bubble, wherein the user does not get to experience items from novel, unseen categories. One approach to mitigate this undesired behavior is to recommend items with high potential for serendipity, namely surprising items that are likely to be highly rated. In this paper, we propose a content-based formulation of serendipity that is rooted in Bayesian surprise and use it to measure the serendipity of items after they are consumed and rated by the user. When coupled with a collaborative-filtering component that identifies similar users, this enables recommending items with high potential for serendipity. To facilitate the evaluation of topic-level models for surprise and serendipity, we introduce a dataset of book reading histories extracted from Goodreads, containing over 26 thousand users and close to 1.3 million books, where we manually annotate 449 books read by 4 users in terms of their time-dependent, topic-level surprise. Experimental evaluations show that models that use Bayesian surprise correlate much better with the manual annotations of topic-level surprise than distance-based heuristics, and also obtain better serendipitous item recommendation performance.
\end{abstract}



\maketitle

\section{Introduction}
\label{sec:introduction}

Recommender systems offer personalized, ranked lists of user-relevant items, easing the cognitive overload caused by the ever growing number of items available to users. Adapting techniques from information retrieval, machine learning, and data mining, recommender systems have a long and rich history, ranging from classical approaches \cite{resnick:acm97,park:esa12} to modern methods based on advances in deep learning and dense representations of users and items \cite{he:www17,zhang:acm19}. Traditionally, recommender systems optimize their output using content-based (CB) signals and collaborative filtering (CF) information, either separately or in combination \cite{melville:aaai02}. Content-based approaches recommend items that are similar to previously consumed liked items, whereas collaborative filtering methods seek to find items that were liked by similar users. By constantly optimizing for these two objectives, recommender systems can create {\it filter bubbles} \cite{pariser_filter_2011,flaxman_filter_2016} where a user is insulated from topics or points of view that are different from the ones they have already been exposed to. This phenomenon is further exacerbated by the feedback loop between the item ranking model and the passive user reaction to the recommended items \cite{masoud:cikm20}, where the $\langle$item, rating$\rangle$ pair is used as a new sample to update the ranking model, thus reinforcing historical user preferences. A widely used concept for alleviating filter bubbles is that of diversity, where recommender systems are encouraged to generate recommendation sets containing dissimilar items \cite{cheng:www17}. Increasing diversity can be done by post-hoc reranking of recommended items, using approaches such as maximal marginal relevance \cite{carbonell:sigir98,ziegler:www05}. Additionally, diversity measures can be incorporated alongside relevance criteria in the objective function and used during training and inference \cite{zheng:www21,wang:sigir22,gao:sigir22}, inter alia. All these methods alleviate the filter bubble by seeking to model and optimize for diversity directly. An alternative strategy is to optimize for measures that have an indirect, but strong positive effect on diversity. Such a measure is serendipity, where the aim is to recommend items that surprise the user in a positive way \cite{adamopoulos2014unexpectedness,adamopoulos2014bias,kotkov2018investigating,niu2018surprise,pandey2018recommending,chen2019serendipity,Li2020serendipity,li2020purs,shrivastava2022optimized}. In this paper, we introduce a formal characterization of topic-level serendipity based on Bayesian surprise, and evaluate its various implementations on a new dataset of book reviews that have been manually annotated with surprise labels.

Reflecting a definition of serendipity as the occurrence of an event that is both {\it surprising} and {\it valuable}, in Sections~\ref{sec:task} and~\ref{sec:online} we introduce a content-based, post-factum measure of serendipity that uses Bayesian surprise \cite{itti_bayesian_2005} and user ratings to capture the two definitional aspects of serendipity. However, by itself, the serendipity component can estimate serendipity only after an item has been consumed and rated. Subsequently, recommendation of serendipitous items is enabled by integrating the measurement of topic-level serendipity with a collaborative filtering component that is tasked with identifying similar users who consumed serendipitous items in the past. Computing Bayesian surprise requires maintaining a probability distribution over user preferences and updating it every time an item is consumed and rated. Correspondingly, in Section~\ref{sec:online} we describe a number of online learning algorithms that were adapted to work in a non-stationary setting where user preferences can drift over time. These are supplemented in Section~\ref{sec:basic} with methods that estimate surprise using distances between point estimates of preference vectors at consecutive times steps in the user's time series of consumed items. For evaluation, in Section~\ref{sec:dataset} we introduce a dataset derived from Goodreads \cite{wan2018goodreads} in which users are associated time series of books and ratings, and where books are manually annotated as to whether they present topic-level surprise at the time they appeared in the user's time series. The experiments in Section~\ref{sec:experiments} show that methods that rely on Bayesian surprise are better at predicting topic-level serendipity. 

\section{Task Definition}
\label{sec:task}

We assume there is a set of books, or more generally a set of {\bf items} $\mathcal{I}$. There is also a set of $K$ {\bf topics}, and for each book item $i \in \mathcal{I}$ there is a corresponding {\bf topic distribution} $\bm{\theta}_i = [\theta_{i,1}, \theta_{i,2}, ..., \theta_{i,K}]$, where $\sum_k \theta_{i,k} = 1$. Furthermore, 
there is a set of {\bf users} $\mathcal{U}$ who consume items over time. For a given user $u \in \mathcal{U}$, let the vector $\mathbf{p}_t(u) = [p_{t, 1}, p_{t, 2}, ..., p_{t, K}]$ capture his set of {\bf user preferences} over the $K$ topics at time $t$, where $p_{t, k}$ represents the user's preference for topic $k$ at time $t$ (the argument $u$ was left implicit). User preferences are unbounded and can be positive or negative, corresponding to the user liking or disliking that topic, respectively. At a time step $t$, upon consuming item $\bm{\theta}_i$, the user experiences a {\bf reward} $r_t$, which expresses how much he\footnote{The user gender was sampled at random by tossing a coin.} liked (positive reward) or disliked (negative reward) that item. Henceforth, to simplify notation, we use $\bm{\theta}_i$ to denote either item number $i$ or the item consumed by the user at time step $i$, where the user is evident from the context. To recommend items with high serendipity, we rely on the following two components:
\begin{enumerate}
    \item {\bf Item Surprise}: Estimate the amount of surprise $Sur(u, t)$ that the book $\theta_t$ generated for a user $u$ who has consumed and rated the books $\langle \mathbf{\theta}_1, r_1 \rangle$, ..., $\langle \mathbf{\theta}_{t-1}, r_{t-1} \rangle$, and $\langle \mathbf{\theta}_t, r_t \rangle$, in this order.
    \item {\bf User Similarity}: Given user $u$ at time step $i$ and another user $v$ at time step $j$, estimate how different the two users are in terms of their preferences, as a distance $d(\langle u, i\rangle, \langle v, j \rangle)$.
\end{enumerate}
Given the two components above, recommending serendipitous items to a user $u$ at a time step $i + 1$ will be done in a collaborative filtering manner by identifying users $v$ who at a time step $j$ in their past are similar in terms of their preferences to user $u$ and furthermore who, at their next time step $j + 1$, consumed an item that was {\it positively} rated and that also resulted in a high level of {\it surprise}, hence {\it serendipitous}. This recommendation procedure is detailed in Algorithm~\ref{alg:ser-item}, as follows. In line 1, the algorithm searches in the entire database of users $v$ and their time series of consumed items $j$ to find the $N$ users whose preferences upon consuming item $j$ were most similar with user $u$'s preferences upon consuming item $i$. This set is further filtered in line 2 to keep only those users whose similarity is not below a predefined threshold and who rated positively the next item. Of these users, line 3 identifies the one whose next item resulted in the largest amount of surprise, which is returned in line 4.

\LinesNumbered
\begin{algorithm}[t]
\caption{\textsc{FindSerendipity}($u$, $i$)}
\label{alg:ser-item}

\KwIn{A reference user $u$ and a time step $i$.}
\nonl\myinput{The preference distance threshold $\tau_d$ (hyperparameter).}
\nonl\myinput{The number $N$ of most similar users to consider (hyperparameter).}
\KwOut{A pair $\langle v, j \rangle$ similar to $\langle u, i \rangle$ where item $j+1$ has high serendipity for user $v$.}
\vspace{1em}

Let $T$ = the top set of $N$ pairs $\langle v, j \rangle$ most similar to $\langle u, i\rangle$ \tcc*{$N$ lowest $d(\langle u, i\rangle, \langle v, j \rangle)$}
Let $S = \{\langle v, j \rangle \in T \mid d(\langle u, i\rangle, \langle v, j \rangle) < \tau_d\ \wedge r_{j+1} > 0 \}$   \tcc*{high similarity and positive rating}
Let $\langle v, j\rangle = \displaystyle\argmax_{\langle v, j\rangle \in S, \; v \neq u} Sur(v, j + 1)$ \tcc*{next item with maximum surprise}
\KwRet{$\langle v, j\rangle$} \tcc*{return null pair $\emptyset$ if $S$ is empty}
\end{algorithm}

To instantiate the algorithm above, we need to specify how to compute the topic-level surprise measure $Sur(u, t)$ and the user preference distance $d(\langle u, i\rangle, \langle v, j \rangle)$. To estimate the surprise $Sur(u, t)$ we investigate two types of approaches:
\begin{itemize}
    \item {\bf Online learning}: In this approach (Section~\ref{sec:online}), the preference vector $\mathbf{p}_t(u)$ is updated at every step to minimize the rating loss $(\hat{r}_t - r_t)^2$ incurred from predicting a rating $\hat{r}_t$ at that time step. We consider linear reward estimation models $\hat{r}_t = \mathbf{p}^T \bm{\theta}_{t}$, where $\mathbf{p}$ can be seen as the parameters and $\bm{\theta}_{t}$ the input features at time $t$. Bayesian surprise $Sur(u, \bm{\theta}_t)$ will then be calculated as the KL divergence between the preference distributions at times $t$ and $t-1$, whereas 
    \item {\bf Basic model}: In this approach (Section~\ref{sec:basic}), simple weighted averages of topic distributions and heuristic distances to compute preference vectors $\mathbf{p}_t(u)$ and surprise values $Sur(u, t)$, respectively.
\end{itemize}
In all approaches, the preference distance $d(\langle u, i\rangle, \langle v, j \rangle)$ will be computed simply as the L2 norm of the difference between the two preference vectors $\mathbf{p}_i(u)$ and $\mathbf{p}_j(v)$, i.e., $d(\langle u, i\rangle, \langle v, j \rangle) = \norm{\mathbf{p}_i(u) - \mathbf{p}_j(v)}_2$.

In the Goodreads dataset, readers provide ratings that are on a scale of 1 to 5 stars. These ratings are projected onto the $[-2, 2]$ interval by subtracting 3 stars from the raw star values (henceforth, we use the terms reward and rating interchangeably). Using the linear reward estimation model, training sequential learning models to fit a label in $[-2, 2]$ is approached as an online linear regression (LR) task. In this context, we define the post-factum {\bf serendipity} of item $\theta_t$ for user $u$ as the product between the rating $r_t$ and the surprise $Sur(u, t)$ that $\theta_t$ triggered for $u$:
\begin{equation}
    Ser(u, t) = r_t \times Sur(u, t)
\end{equation}
Given that all surprise measures described in Section~\ref{sec:online} below are positive and a positive reward corresponds to a raw rating strictly greater than 3 stars, the definition above maps well to the conceptual definition of serendipity as an event that is both surprising, i.e., large $Sur(u, t)$, and valuable, i.e., $r_t > 0$. Looking back at Algorithm~\ref{alg:ser-item}, by only selecting items $j + 1$ with positive rating (line 3) that maximize surprise (line 4), the procedure can be seen as returning an item with high serendipity value for $v$. Since $v$ itself was selected to have similar preferences with user $u$, the collaborative filtering expectation is that item $j + 1$ will be serendipitous for user $u$ as well.


\section{Online Learning of Non-Stationary User Preferences}
\label{sec:online}

We assume that at each time step $t$ a user's preference vector has a multivariate normal distribution $\mathbf{p}_t \sim \mathcal{N}(\bm{\mu}_t, \Sigma_t)$. In this section, we describe online learning algorithms that take as input the preference distribution at the previous time step $\mathbf{p}_{t-1} \sim \mathcal{N}(\bm{\mu}_{t-1}, \Sigma_{t-1})$, the topic vector for the current item $\bm{\theta}_t$, and the user's rating $r_t$, and update the preference distribution to $\mathbf{p}_t \sim \mathcal{N}(\bm{\mu}_t, \Sigma_t)$ in order to minimize the squared distance between the predicted rating $\hat{r}_t$ and the true rating $r_t$. Under a Bayesian interpretation, $\mathbf{p}_{t-1}$ is the prior preference distribution whereas $\mathbf{p}_t$ is the posterior distribution upon observing item $\bm{\theta}_t$ and rating $r_t$. Bayes rules is indeed how the posterior distribution is derived in the Bayesian linear regression approach described in Section~\ref{sec:vblr} below.

Under the framework of Bayesian surprise of \citet{itti_bayesian_2005}, the {\bf surprise} elicited by the item $\bm{\theta}_t$ from user $u$ is defined as the distance between the posterior and prior distributions, measured using KL divergence \cite{kullback_information_1959}:
\begin{equation}
   Sur(u, t) =  KL\left(\mathcal{N}(\mu_t, \Sigma_t) || \mathcal{N}(\mu_{t-1}, \Sigma_{t-1})\right)
\end{equation}
In Sections~\ref{sec:vblr} and~\ref{sec:arow} we introduce adaptations of Bayesian linear regression and adaptive regularization of weights (AROW) for the estimation of non-stationary user preference distributions, which will enable calculating Bayesian surprise as defined above. Note that these methods will require the rating $r_t$ to compute the posterior distribution, which means surprise will be measured post-factum, and may be caused by both novel topic distributions and unexpected ratings (a different definition using only the covariance matrices is left for future work). A post-factum definition of surprise is also used in the NLMS approach from Section~\ref{sec:nlms}. 

\subsection{Variance Bounded Bayesian Linear Regression}
\label{sec:vblr}

The first sequential learning model that we consider for updating the user preference distribution is that of Bayesian linear regression \cite{bishop_pattern_2013}. Under a Bayesian treatment of linear regression, at the beginning $t = 0$ of a user's reading history, its preference vector is distributed according to a zero-centered Gaussian prior $\mathbf{p}_0 \sim \mathcal{N}(\bm{0}, \beta I)$. Due to the choice of a conjugate Gaussian prior distribution, it can be shown that the posterior at the next time step will be Gaussian as well according to the following update rules (section 3.3 in \cite{bishop_pattern_2013}):
\begin{equation}
   \Sigma_{t+1}^{-1} = \Sigma_t^{-1} + \beta \bm{\theta}_t^T \bm{\theta}_t \mbox{\ \ \ and \ \ \ } \bm{\mu}_{t+1} = \Sigma_{t+1} \left(\Sigma_t^{-1} \bm{\mu}_t + \beta \bm{\theta}_t^T r_t \right) \label{eq:blr}
\end{equation}

The original Bayesian LR model is intended for a {\it stationary} setting where the true model that generates the data and their labels does not change over time. However, in a recommendation setting, users change their preferences over time, which requires the LR model to be flexible enough to allow for sometimes drastic changes in its parameters, depending on how much the true user preferences changed. Using the original Bayesian LR model in such a {\it non-stationary} setting is then going to be suboptimal due to the fact that once the parameter co-variance $\Sigma$ gets close to 0, the parameters change very little. To alleviate this issue, we introduce {\it variance bounded Bayesian LR} that ensures the variance of each parameter is at least $\tau_v$, where $\tau_v > 0$ is a hyperparameter. At every update step, we take the current covariance matrix $\Sigma_t$, perform an eigenvalue decomposition $\Sigma_t = U_t \Lambda U_t^T$, where $\Lambda = diag(\lambda_1, ..., \lambda_K)$ is the diagonal matrix of positive eigenvalues. Then all eigenvalues are clipped to be at least $\tau_v$ and used in a new diagonal matrix of clipped eigenvalues $\Lambda_v = clip(\Lambda, \tau_v) = diag(max(\lambda_1, \tau_v), ..., max(\lambda_K, \tau_v))$. We use $\Lambda_v$ to compute a new covariance matrix $S_t = U_t \Lambda_v U_t^T$, which is then employed as usual in the Bayesian LR update equations, as shown below.
\begin{eqnarray}
    \Sigma_t = U_t \Lambda U_t^T \mbox{\ \ } \longrightarrow \mbox{\ \ } 
    \Lambda_v =  clip(\Lambda, \tau_v) \mbox{\ \ } \longrightarrow \mbox{\ \ }
    S_t  =  U_t \Lambda_v U_t^T \\
    \Sigma_{t+1}^{-1} = S_t^{-1} + \beta \bm{\theta}_t^T \bm{\theta}_t  \mbox{\ \ \ and \ \ \ }  \bm{\mu}_{t+1} = \Sigma_{t+1} \left(S_t^{-1} \bm{\mu}_t + \beta \bm{\theta}_t^T r_t \right)  
\label{eq:vblr-}
\end{eqnarray}


\subsection{Adaptive Regularization of Weights for Regression}
\label{sec:arow}


The Adaptive Regularization of Weights (AROW) algorithm \cite{crammer_adaptive_2009} is an online optimization procedure for confidence-weighted learning of linear classifiers \cite{dredze2008confidence}. The original optimization formulation of AROW is:
\begin{equation}
    \mu_t, \Sigma_t = \argmin_{\mu, \Sigma} C(\mu,\Sigma) \mbox{,\ \ \ where } C(\mu,\Sigma) = KL\left(\mathcal{N}(\mu, \Sigma) || \mathcal{N}(\mu_{t-1}, \Sigma_{t-1})\right) + \frac{1}{2 r_1} l(r_t, \mu^T \bm{\theta}_t) + \frac{1}{2 r_2} \bm{\theta}_t^T \Sigma \bm{\theta}_t
\end{equation}
where $r_1$ and $r_2$ are two hyperparameters quantifying the trade-off between the prediction accuracy and the confidence in the new parameters. By minimizing the first term (Bayesian surprise), the new parameters are encouraged to stay close to the current values. In the original AROW, the loss $l$ was set to be the squared hinge loss for classification.

We adapt AROW for regression by replacing the original squared-hinge loss with the squared error loss $l(r_t, \mu^T \theta_t) = (r_t - \mu^T \theta_t)^2$. Furthermore, whereas the hyperparameters $r_1$ and $r_2$ were set to be equal in \cite{crammer_adaptive_2009}, in our experiments they are tuned separately. In a non-stationary setting it is especially important that $r_2$ is independent from $r_1$, to allow the parameters to vary more freely when the user changes his preferences.

To solve for the parameters $\mu, \Sigma$ that minimize equation (1), we use a derivation analogous to \cite{crammer_adaptive_2009} by writing the KL term explicitly and decomposing the loss in two parts depending on $\mu$ and $\Sigma$, i.e., $C(\mu,\Sigma) = C_1(\mu) + C_2(\Sigma)$:
\begin{eqnarray}
    C_1(\mu) & = & \frac{1}{2}(\mu_{t-1}-\mu)^T \Sigma_{t-1}^{-1}(\mu_{t-1}-\mu) +\frac{1}{2r_1} (r_t - \mu_t^T \bm{\theta}_t)^2 \label{eq:arow-mu} \\
    C_2(\Sigma) & = & \frac{1}{2}\log{\frac{det\,\Sigma_{t-1}}{det\,\Sigma}} + \frac{1}{2}Tr(\Sigma_{t-1}^{-1}\Sigma) + \frac{1}{2r_2} \bm{\theta}_t^T\Sigma \bm{\theta}_t \label{eq:arow-sigma} 
\end{eqnarray}
By settings the gradients of Equations~\ref{eq:arow-mu} and~\ref{eq:arow-sigma} to zero, we obtain the following online update equations:
\begin{equation}
    \mu_t = \mu_{t-1} + \frac{(r_t - \mu_{t-1}^T \bm{\theta}_t)\Sigma_{t-1}\bm{\theta}_t}{r_1 + \bm{\theta}_t^T\Sigma_{t-1}\bm{\theta}_t} \mbox{\ \ \ \ \ and \ \ \ \ \ }
    \Sigma_t = \Sigma_{t-1} - \frac{\Sigma_{t-1}\bm{\theta}_t\bm{\theta}_t^T\Sigma_{t-1}}{r_2 + \bm{\theta}_t^{T}\Sigma_{t-1}\bm{\theta}_t}
\end{equation}
Note that by keeping $r_1$ and $r_2$ separate, this formulation is different from the AROW version in \cite{vaits_re-adapting_2011}. 

\subsection{Normalized Least Mean Square}
\label{sec:nlms}

We also experiment with normalized least mean square (NLMS) \cite{bershad1986analysis}, a more stable version of the well known least mean square (LMS) algorithm for regression, where the learning rate is divided by the squared norm of the feature vector:
\begin{equation}
    \mbox{Preference } \bm{p}_t = \bm{p}_{t-1} + \frac{\eta}{\bm{\theta_t}^T \bm{\theta_t}} (r_t - \bm{p}_{t-1}^T \bm{\theta_t}) \bm{\theta_t} \mbox{\ \ \ and \ \  Surprise } Sur(v, t) = \norm{\bm{p}_{t + k - 1} - \bm{p}_{t - 1}}
\end{equation}
Note that NLMS only updates a point estimate $\bm{p}_t = \bm{\mu}_t$ of the preference vector, therefore it cannot be used to compute Bayesian surprise. Instead, we define the surprise at time $t$ as the the norm of the difference between the preference vectors at times $t - 1$ and $t + k - 1$, where $k \geq 1$ is a time horizon hyper-parameter.
When the time horizon is given the default value $k = 1$, this measure can be seen as capturing the impact that the item $\theta_t$ had on the preference vector update. Larger values of $k$ are meant to model the fact that the impact of an item item $\theta_t$ may become clearer only after multiple NLMS updates. While it may appear that preference vectors from the "future" are used, this future is only relative to a point $t$ in the past for a user $v$ and it is not an issue as long as it is done only for finding similar users $v$ and their post-factum surprising items in the collaborative filtering step from Algorithm~\ref{alg:ser-item} (no future information is used for the reference users $u$ during evaluation).


\section{Basic Model for Surprise and User Preferences}
\label{sec:basic}

A simpler approach is to decouple the rating from surprise and define surprise as a distance $d$ between an item $\bm{\theta}_t$ and the previously consumed items summarized in a topic history vector $\bm{h}_{t-1}$:
\begin{equation}
    Sur(u, t) = d(\bm{\theta}_t, \bm{h}_{t-1}) \mbox{\ \ \ \ where \ \ \ \ } \bm{h}_{t-1} = \frac{1}{t-1} \sum_{z=1}^{t-1} \bm{\theta}_{z}
\end{equation}
Here, $\bm{h}_{t-1}$ embeds the history of items into an average topic distribution vector aimed at expressing the topics that the user has consumed so far. We also tried using an exponential decay hyper-parameter that is meant to give lower weight to items seen farther away in the past, and thus better accommodate a non-stationary setting, however the best results were obtained with the simpler, and perhaps more stable, topic average.

In terms of the actual distance function $d$, we discovered that the Euclidean distance $d(\bm{\theta}_t, \bm{h}_{t-1}) = \norm{\bm{\theta}_t - \bm{h}_{t-1}}$ did poorly at identifying surprising items because it was often dominated by many topics that were in common between different items, even when they had small probabilities. Instead, to calculate surprise we use the topic that stands out the most with respect to the maximum topic probability so far:
\begin{equation}
    d(\bm{\theta}_t, \bm{h}_{t-1}) = \norm{\bm{\theta}_t - \bm{m}_{t-1}}_{\infty} = \max_{1 \leq k \leq K} \left(\theta_{t,k} - m_{t-1,k}\right) \mbox{\ \ \ \ where \ \ \ \ } m_{t-1,k} = \max_{1 \leq z \leq t - 1} h_{z, k}
\end{equation}
The vector $\bm{m}_{t-1} = [m_{t-1,k}]_{k = 1..K}$ stores for each topic $k$ the largest probability with which it was seen across all the books read so far by the user $u$.

The preference vector is computed in the basic model as the weighted average of the topic distributions seen by the user, using their 1 to 5 ratings as weights: 
\begin{equation}
    \bm{p}_t = \frac{1}{t-1} \sum_{z=1}^{t-1} r_{z} \bm{\theta}_{z}
\end{equation}
Here too we tried using an exponentially decaying weights for past items, however best performance was obtained using the simple weighted average.

\section{Topic-Level Surprise and Serendipity Dataset}
\label{sec:dataset}

We use as raw data source the Goodreads dataset \cite{wan2018goodreads}, which consists of over 15M reviews for about 2M books from around 465K users. For lack of access to the actual book contents, we associate each book ID with an artificial {\it book content} that is created by concatenating all its English reviews and trimming to a maximum of 10K tokens. We extract a main set of 1,294,532 {\sc Books} and their reviews by considering only books with at least 50 tokens. We create a smaller subset {\sc LDABooks} for training an LDA topic model using only the 303,832 books that contained at least 1,000 tokens in their book content. Furthermore, we extracted a main set of 26,374 {\sc Users} by considering only those who read and rated at least 100 different items from the {\sc Books} set. While this may lead to selection bias, it was done to minimize the chance of including users with missing book ratings, based on the assumption that users who submit many ratings are more likely to report every book they read on Goodreads. Overall, the reading histories of these {\sc Users} contain 1,043,437 unique books.

\subsection{Manual Annotation of Topic-Level Surprise}

To enable evaluation of the various serendipity models introduced above, we sampled a set of 4 {\it reference users} with reading histories that appeared to be diverse, in order to increase the likelihood of finding books that are topic-level surprising. For every book in a user's reading history, we read through the book content (concatenated reviews) in order to (a) manually create a {\it list of the major topics} for that book; and (b) use that list and the book content to create a {\it short summary} of the book. Sometimes the accumulated book content was too small to get a clear idea of the book topics, as such we queried Goodreads and Amazon for additional reviews that were then added to the book content field.

For each of the 4 reference users, we manually annotate the books in their reading history with binary surprise labels. We skip the first 15 books, as these will be used as a burn-in set to learn a more stable estimate of the user's initial preference vector. Starting with position 16 in the time series of books, we compare its list of major topics and the book summary with a running summary of the topics of the books read so far, as well as their summaries. If the current book has topics that have not appeared in the previous books, or that were only tangentially addressed in previous books, then the book is labeled as topic-level surprising. Note that topic-level surprise is only one of many types of surprise, some of which are even independent of content, e.g. the user discovering a book by accident \cite{fu:sigir23}. Table~\ref{tab:dataset} shows annotation statistics for each reference user, including the number of surprising books.
Overall, this is a time consuming, cognitively demanding annotation exercise. Although the total number of books that are manually labeled for surprise is not small, the requirement and difficulty of annotating each book with respect to the previously read books limited the annotation of book reading histories to only 4 users. To facilitate reproducibility and future progress in this area, we make the code and the dataset publicly available at \href{https://github.com/Ton-moy/surprise-and-serendipity}{https://github.com/Ton-moy/surprise-and-serendipity}. 

\begin{table}[t]
    \centering
    \caption{Reference users statistics. The total number of books for each user is 15 more than the number of manually labeled books.}
    \begin{tabular}{|l|cccc|c|}
    \cline{2-6}
       \multicolumn{1}{c|}{} &  {User 1}  &  {User 2}  & {User 3} & {User 4} & Total\\ \hline
       Books read  &  {140}  &  {114} & {137} & {118} & 509 \\ \hline
       Manually labeled books  &  {125}  &  {99} & {122} & {103} & 449 \\ \hline
       Surprising books  &  {10}  &  {13} &  {14} & {17} & 54 \\ \hline
    \end{tabular}
    \label{tab:dataset}
\end{table}



\section{Experimental Evaluation}
\label{sec:experiments}

We train a Latent Dirichlet Allocation (LDA) \cite{blei2003lda} topic model with $K = 100$ topics on the {\sc LDABooks} set, and use it to generate the input topic distributions for all {\sc Books}. We evaluate a total of 6 models: (1) Bayesian linear regression BLR($\beta$, $\tau_s$, $\tau_d$, $N$); (2) variance bounded Bayesian linear regression vbBLR($\beta$, $\tau_v$, $\tau_s$, $\tau_d$, $N$); (3) AROW regression AROW($r_1$, $r_2$, $\tau_s$, $\tau_d$, $N$); (4) normalized least mean square NLMS($\eta$, $\tau_s$, $\tau_d$, $k$, $N$), (5) the hybrid combination AROW+vbBLR; and (6) the basic model Basic($\tau_s$, $\tau_d$, $N$). The hyperparameters for each model are indicated between parentheses and are tuned using a {\it leave-one-out} setup: if reference user $u \in U$ is the current test user, then we select the hyperparameter values that lead to the best average $F_1$ on the other 3 users in $U - \{u\}$. This is repeated 4 times, in order to get test results on all reference users. For each method, user preference vectors $\bm{p}_t(u)$ are created for every user $u$ in {\sc Users} at every time step $t$ in their time series of books. To compute precision and recall for the task of serendipity recommendation, we use the procedure shown in Algorithm~\ref{alg:eval} for each reference user. 

\SetAlgoNoLine
\LinesNumbered
\begin{algorithm}[t]
\caption{\sc{EvaluateSerendipityForUser}$(u)$}
\label{alg:eval}
\KwIn{A reference user $u$; $\tau_s$ is the surprise threshold.}
\nonl\myinput{$mSur(u, i + 1)$ is true iff item $i + 1$ was manually labeled as surprising for $u$.}
\KwOut{The Precision ($P$), Recall ($R$), and F1 measure ($F_1$) for reference user $u$.}
\vspace{0.5em}

$tp, tn, fp, fn \gets 0$\;
\For{each item $i$ in user $u$'s time series} {
    \If{$\langle v, j \rangle = $ {\sc FindSerendipity}$(u, i)$ is not null} {
        \eIf{$Sur(v, j+1) > \tau_s$} 
            {
            {\bf if} {$mSur(u, i + 1)$ and $r_{i+1} > 0$} {\bf then} {$tp \gets tp + 1$} {\bf else} {$fp \gets fp + 1$}
            }{
            {\bf if} {$mSur(u, i + 1)$ and $r_{i+1} > 0$} {\bf then} {$fn \gets fn + 1$} {\bf else} {$tn \gets tn + 1$}
            }
    }
}
\KwRet{$P = tp / (tp + fp)$, $R = tp / (tp + fn)$, $F_1 = 2PR / (P + R)$}
\end{algorithm}

The surprise detection results are listed in Table~\ref{tab:surprise} and show the AROW model outperforming all other models, achieving an average F1 of 56.5\%. The Basic model has the lowest average F1 of 42.7\%. The results also show that models that utilize Bayesian surprise outperform other models by a considerable margin. The last two rows show the random baseline performance using either a surprise probability of 0.5, or the ratio of surprising books observed in the data. The serendipity recommendation performance is reported in Table~\ref{tab:serendipity}. Because AROW obtained the best surprise recommendation performance and vbBLR the best serendipity recommendation, we also evaluated a hybrid AROW+vbBLR combination, where AROW is used for computing surprise, and vbBLR is used for identifying the most similar users in Algorithm~\ref{alg:ser-item}. This combination outperforms all other models, obtaining an average F1 of 37.0\%. The Basic model's performance is very unstable; while it obtains competitive performance on the first 3 users, on user 4 all the book items that it recommends as serendipitous are wrong, leading to no true positives and consequently zero F1. A possible reason is because the hyperparameters that obtain the best performance when tuned on the first 3 users are not suitable for user 4. However, even tuning on the user 4 itself lead to a very low F1 of around 13\%.

Error analysis for the best models reveals that most errors are caused by (a) reviewers writing about topics that are unrelated to the book content, such as movies made based on the book, or books written by the same author; and (b) LDA creating irrelevant topics based on character names and other proper names.

\section{Serendipity vs. other Beyond-Accuracy Recommendation Metrics}

\citet{mcnee_being_2006} observed that focusing solely on the accurate ranking of items that are known or expected by users misses other important aspects that can further amplify user satisfaction. Currently, the recommender systems literature (\cite{ge:acm10,kaminkas:acm16,li:acm20}, {\it inter alia}) recognizes five major non-accuracy aspects: serendipity, unexpectedness, novelty, diversity, and coverage. Diversity refers to how dissimilar generated recommendations are, whereas coverage reflects the degree to which they cover the entire spectrum of available items. The remaining three concepts, namely serendipity, unexpectedness, and novelty, have been variously described using multiple definitions that often result in substantial overlap, subsumption, and sometimes even identity, between their conceptual domains. For example, noting that there is no consensus on the definition of serendipity, \citet{kotkov2018investigating} investigate eight definitions, starting from a common view where serendipity has three components -- relevance, novelty and unexpectedness, each of which has multiple variations. Novelty typically refers to a user being unfamiliar with a recommended item \cite{oh_novel_2011}, a desirable property that is lacking when recommendation lists contain only items that are popular or well-known \cite{Herlocker:acm04}. At the same time, \citet{oh_novel_2011} note that sometimes novel recommendations are equated with diversified recommendations, whereas other approaches define novel items more broadly as any item that widens a user's interests. Definitional ambiguity aside, other than serendipity, unexpectedness is most related to our notion of surprise. \citet{li:acm20} define the unexpectedness of an item as the distance between that item and the closure of all previously consumed items, computed in a latent embedding space, which is conceptually similar to how surprise is defined in the basic approach from Section~\ref{sec:basic}. A hybrid utility function is then defined as a linear combination of the item's predicted rating and its unexpectedness. In contrast, we define serendipity as a multiplicative combination of the post-factum, actual item rating with its estimated Bayesian surprise, and use collaborative filtering to identify items with high potential for serendipity.

\begin{table}[t]
\setlength\tabcolsep{2.5pt} 
\centering
\caption{Surprising item detection performance (\%) evaluated across 449 manually annotated books from 4 reference users.}
\label{tab:surprise}
\begin{tabular}{c|ccc|ccc|ccc|ccc|c|}
\cline{2-14}
                                          & \multicolumn{3}{c|}{User 1}                                                                              & \multicolumn{3}{c|}{User 2}   & \multicolumn{3}{c|}{User 3}    & \multicolumn{3}{c|}{User 4} 
                                           & \multicolumn{1}{c|}{All}
                                          
                                          \\ \cline{2-14} 
                                          & {P} & {R} & {F1}               & {P} & {R} & {F1}       & {P} & {R} & {F1}      &       {P} & {R} & {F1}      &  Avg F1  \\ \hline
\multicolumn{1}{|c|}{BLR}    &  {30.0}          &  {60.0}       &  {40.0}                    &  {50.0}          &  {69.2}       &  {58.1}           &  {64.3}          &  {64.3}       &  {64.3}      &  {55.6}          &  {58.8}       &  {57.1}        & {54.9}            \\ \hline
\multicolumn{1}{|c|}{vbBLR} & {23.1}  & {60.0}   & {33.3}              & {50.0}      & {69.2}   & {58.1} &  {64.3}      & {64.3}   & {64.3}     & {52.3}      & {64.7}   & {57.9}   & {53.4}            \\ \hline
  \multicolumn{1}{|c|}{AROW}     &  {30.0}          &  {60.0}       &  {40.0}                        &  {61.5}          &  {61.5}       &  {61.5}          &  {64.3}          &  {64.3}       &  {64.3}            &  {62.5}          &  {58.8}       &  {60.0}        &  {\bf 56.5}     \\ \hline
  \multicolumn{1}{|c|}{NLMS}                &  {30.0}          &  {60.0}       &  {40.0}      &  {34.6}          &  {69.2}       &  {46.2}        &  {50.0}          &  {35.7}       &  {41.7}         &  {33.3}          &  {64.7}       &  {44.0}          &  {43.0}           \\ \hline
  \multicolumn{1}{|c|}{Basic}       &  {20.0}          &  {30.0}       &  {24.0}      &  {44.4}          &  {61.5}       &  {51.6}        &  {66.7}          &  {42.9}       &  {52.2}      &  {54.5}          &  {35.3}       &  {42.9}        & {42.7} \\ \hline
    \multicolumn{1}{|c|}{Random ($p = 0.5$)}       &  {8.0}          &  {50.0}       &  {13.8}       &  {13.1}          &  {50.0}       &  {20.8}      &  {11.4}          &  {50.0}       &  {18.6}       &  {16.5}          &  {50.0}       &  {24.8}       & {19.5}  \\ \hline
   \multicolumn{1}{|c|}{Random ($p = P/T$)}       &  {8.0}          &  {8.0}       &  {8.0}       &  {13.1}          &  {13.1}       &  {13.1}        &  {11.4}          &  {11.4}       &  {11.4}        &  {16.5}          &  {16.5}       &  {16.5}        & {12.3}  \\ \hline
\end{tabular}
\end{table}  

\begin{table}[t]
\centering
\setlength\tabcolsep{2.5pt} 
\caption{Serendipitous item recommendation performance (\%), across 449 manually annotated books from 4 reference users.}
\label{tab:serendipity}
\begin{tabular}{c|ccc|ccc|ccc|ccc|c|}
\cline{2-14}
                                          & \multicolumn{3}{c|}{User 1}                                                                              & \multicolumn{3}{c|}{User 2}   & \multicolumn{3}{c|}{User 3}    & \multicolumn{3}{c|}{User 4}   & \multicolumn{1}{c|}{All}                                                                         \\ \cline{2-14} 
                                          & {P} & {R} & {F1}                & {P} & {R} & {F1}      & {P} & {R} & {F1}     &       {P} & {R} & {F1}   &  Avg F1   \\ \hline
\multicolumn{1}{|c|}{BLR}    &  {13.8}  & {57.1}   & {22.2}                       &  {30.8}          &  {80.0}       &  {44.4}        &  {22.7}      & {41.7}   & {29.4}   &  {22.7}      & {50.0}   & {31.3}       &  {31.8}          \\ \hline
\multicolumn{1}{|c|}{vbBLR} & {16.7}  & {71.4}   & {27.0}                  & {30.0}      & {90.0}   & {45.0}      & {20.0}      & {41.7}   & {27.0}     & {30.0}      & {60.0}   & {40.0}       & {\bf 34.8}       \\ \hline
  \multicolumn{1}{|c|}{AROW}     &  {14.3}          &  {57.1}       &  {22.9}                           &  {31.3}          &  {50.0}       &  {38.5}            &  {23.8}          &  {41.7}       &  {30.3}                   &  {22.7}       &  {50.0}     &   {31.3}       &  {30.8}   \\ \hline
  \multicolumn{1}{|c|}{NLMS}                &  {12.1}          &  {57.1}       &  {20.0}      &  {18.8}          &  {33.3}       &  {24.0}      &  {15.4}          &  {36.4}       &  {21.6}      &  {17.9}          &  {50.0}       &  {26.3}            &  {23.0}          \\ \hline
  \multicolumn{1}{|c|}{Basic}       &  {16.1}          &  {71.4}       &  {26.3}       &  {32.0}          &  {72.7}       &  {44.4}      &  {44.4}          &  {33.3}       &  {38.1}      &  {0.00}          &  {0.00}       &  {0.00}       & {27.2}  \\ \hline
   \multicolumn{1}{|c|}{AROW+vbBLR}       &  {19.2}          &  {71.4}       &  {30.3}       &  {31.8}          &  {70.0}       &  {43.8}      &  {22.7}          &  {41.7}       &  {29.4}      &  {35.3}          &  {60.0}       &  {44.4}       & {\bf 37.0}  \\ \hline
   \multicolumn{1}{|c|}{Random ($p = 0.5$)}       &  {5.6}          &  {50.0}       &  {10.1}      &  {11.1}          &  {50.0}       &  {18.2}     &  {9.8}          &  {50.0}       &  {16.4}      &  {10.7}          &  {50.0}       &  {17.6}       & {15.6}  \\ \hline
   \multicolumn{1}{|c|}{Random ($p = P/T$)}       &  {5.6}          &  {5.6}       &  {5.6}       &  {11.1}          &  {11.1}       &  {11.1}      &  {9.8}          &  {9.8}       &  {9.8}      &  {10.7}          &  {10.7}       &  {10.7}       & {9.3}  \\ \hline

\end{tabular}
\end{table}




\section{Conclusion and Future Work}

We introduced a method for recommending serendipitous items where serendipity is defined in terms of Bayesian surprise. To facilitate its computation, we proposed adaptations of online learning algorithms for the non-stationary setting of user preferences. Experiments show that methods rooted in Bayesian surprise obtain superior results. Future work includes an expanded dataset, non-linear models, deep topic modeling, and surprise measures that go beyond topic-level. Since Bayesian linear regression can be obtained as a special case of the Kalman filter, an intriguing future direction is adapting a Kalman filter for continual online learning \cite{titsias2023kalman} of the non-stationary user preferences.

\begin{acks}
We would like to thank Cade Mack for initial implementations of the BLR and vbBLR models and their evaluations on artificial data. We are also grateful to the anonymous reviewers for their constructive comments. This research was partly supported by the United States Air Force (USAF) under Contract No. FA8750-21-C-0075.
Any opinions, findings, conclusions, or recommendations expressed in this material are those of the author(s) and do not necessarily reflect the views of the USAF.
\end{acks}

\bibliographystyle{ACM-Reference-Format}
\bibliography{recsys23}


\end{document}